\documentclass{icbo}
\RequirePackage{filecontents} 
\usepackage{float}
\usepackage{cite}
\usepackage{booktabs}
\usepackage{tabulary}
\usepackage{graphicx}
\usepackage{xspace}
\usepackage[table]{xcolor}
\usepackage{float}
\usepackage{listings}
\usepackage{color}
\usepackage{url}

\newcolumntype{L}{>{\raggedright\arraybackslash}m{10.8cm}}

\begin{document}

\title[Using Literature to Define Biomedical Ontologies' Scope]{A
  Literature Based Approach to Define the Scope of Biomedical
  Ontologies: A Case Study on a Rehabilitation Therapy Ontology}

\author[Halawani \textit{et~al}]{Mohammad
  K. Halawani\,$^1$\,$^,$\,$^3$\footnote{To whom correspondence should
    be addressed: M.K.H.Halawani2@newcastle.ac.uk}, Rob Forsyth\,$^2$
  and Phillip Lord\,$^1$}

\address{$^{1}$School of Computing Science \\
  Newcastle University, UK \\
  $^{2}$Institute of Neuroscience \\
  Newcastle University, UK \\
  $^{3}$Department of Information Systems \\
  Umm Al-Qura University, Saudi Arabia }

\maketitle

\begin{abstract}

  In this article, we investigate our early attempts at building an
  ontology describing rehabilitation therapies following brain
  injury. These therapies are wide-ranging, involving interventions of
  many different kinds. As a result, these therapies are hard to
  describe. As well as restricting actual
  practice, this is also a major impediment to evidence-based medicine
  as it is hard to meaningfully compare two treatment plans.

  Ontology development requires significant effort from both
  ontologists and domain experts. Knowledge elicited from domain
  experts forms the scope of the ontology. The process of knowledge
  elicitation is expensive, consumes experts' time and might have
  biases depending on the selection of the experts. Various
  methodologies and techniques exist for enabling this knowledge
  elicitation, including community groups and open development
  practices. A related problem is that of defining scope. By defining
  the scope, we can decide whether a concept (i.e. term) should be
  represented in the ontology. This is the opposite of knowledge
  elicitation, in the sense that it defines what should not be in the
  ontology. This can be addressed by pre-defining a set of competency
  questions.

  These approaches are, however, expensive and time-consuming. Here,
  we describe our work toward an alternative approach, bootstrapping
  the ontology from an initially small corpus of literature that will
  define the scope of the ontology, expanding this to a set covering
  the domain, then using information extraction to define an initial
  terminology to provide the basis and the competencies for the
  ontology. Here, we discuss four approaches to building a suitable
  corpus that is both sufficiently covering and precise.
  
\end{abstract}

\section{Introduction}
\label{Introduction}

Rehabilitation therapies, unlike pharmacologic therapies, are
difficult to define precisely both qualitatively and
quantitatively~\citep{van2012evidence} and many approaches have been
taken to trying to parse them. It is recognised that traditional
approaches to designation (e.g. ``dressing practice'') are flawed as
two professionals' rehabilitation sessions both targeting difficulties
in dressing could differ in pertinent active ingredients
(e.g. actions, chemicals, devices, or forms of energy)
as experienced by the patient. Assumptions that rehabilitation content
can be inferred from the targeted impairment (e.g. ``balance
training'') as flawed: no one would consider it appropriate to
consider bariatric surgery, calorie-restricted diets and exercise
programmes together as equivalent forms of ``obesity
therapy''~\citep{whyte2014development}. This lack of a shared
terminology makes it difficult to describe, measure and meaningfully
compare rehabilitation therapies and treatments.

Building a taxonomy for rehabilitation treatments could lead to a
better shared understanding of rehabilitation
interventions~\citep{dijkers2014rehabilitation}. Hence, a
rehabilitation treatment ontology (RTO) of rehabilitation terms, as
the terms represent the concepts and knowledge of the
domain~\citep{sowa2000ontology}, should ease the dissemination of
treatments to communicate about them clearly and effectively, through
a shared understanding.

To enable building the RTO, we need to define both the terms that we
wish to be in the ontology and those that should not. Some ontologies
have extremely well-defined scopes, such as the Karyotype
ontology~\citep{warrender2013karyotype}, which is an ontological
representation of a previously defined informal specification. Others,
such as the mitochondrial disease
ontology~\citep{warrender2015consistent} relate to a specific area of
knowledge, or like the Gene Ontology(GO)~\citep{ashburner2000gene} to
a broad area, but at a specific granularity. For the RTO,
unfortunately, the breadth of the area means that we lack this clear
statement of scope.

Of course, there has been significant research on \emph{ontology
  learning}, enabling either automation or semi-automation of the
ontology construction process~\citep{buitelaar2005ontology}. For the
RTO, we aim to use a semi-automated approach, combined with a highly
programmatic, pattern-driven ontology construction methodology that we
have pioneered previously with the mitochondrial disease
ontology~\citep{warrender2015scaffolding}: this separates terms out
into a \textit{scaffold} generated automatically, often from a
pre-existing structured source such as a database. This is followed by
manual refinement using the vocabulary provided this scaffold.

With the RTO, we plan to extend this ontology construction
methodology: first, we will build a corpus of appropriate literature
that will define the scope of the ontology; then we can use
this to extract a set of representative terms and phrases; finally, we
will use these terms and phrases as the basis for our
\textit{ontological scaffold}~\citep{warrender2015scaffolding}. This
should provide both coverage and scope for our ontology, which we can
then refine and build further either manually or through the addition
of further scaffolded terms, identified during the first phase of
development. We have previously used a similar methodology to ensure
good coverage and define the scope of MITAP, a minimum information
model~\citep{lord2016minimum}.

This leaves us with the problem of defining an appropriate corpus of
literature for the RTO. This corpus needs to cover the domain
adequately; at the same time, we would like this corpus to be
reflective of opinions of a wider community than the experts involved
it its construction. This is a common problem with ontology
development: if the scope is too narrow, the ontology will fulfil the
needs of only a few; if it is too broad, the ontology will either get
large or only have general terms. 

The aim of this article is to investigate different semi-automated
methods and search strategies to retrieve a corpus with a high level
of accuracy and coverage with respect to the communities needs for the
RTO. The accuracy and coverage of a corpus are its precision and
recall, respectively, in relation to the scope of rehabilitation. We
describe four different techniques that we have used all based around
use of PubMed, and describe their advantages and disadvantages.

\section{Methods}
\label{sec:methods}

For this work, we have used PubMed exclusively to define our
corpus. As a corpus, PubMed is far from ideal. While it contains many
papers about rehabilitation, they are mostly written from an academic
perspective and may make a different use of vocabulary from the
clinicians. A significant percentage of the papers
in PubMed have only abstracts accessible (although, under UK law, we
may be able to access full text by other
means~\citep{greycite81781}). However, it has other significant
advantages: it is freely available; there are no patient
confidentiality restrictions as there would be with medical records;
finally, it has a good API and is easy to access computationally.

We use two additional features of PubMed in this paper. First, papers
are annotated with Medical Subject Headings (MeSH). MeSH is a
thesaurus organised into a hierarchy; searches with a single term,
also search the transitive closure of that term. Curators can also
define a MeSH annotation as the ``major term'' or MAJR. Secondly,
PubMed provides a similar articles functionality (PMSA), based on text
similarity~\citep{pub2001similar}. Currently, this functionality only
allows retrieving MEDLINE records (i.e. PubMed citation) similar
to a single user-selected record. We discuss this limitation later.

Additional search functionality described in this paper was
implemented using Python, exploiting the Entrez module of
BioPython~\citep{cock2009biopython}.

\subsection{Forming a Corpus}
\label{sec:forming-corpus}

The simplest approach to generating a suitable corpus is a
\textbf{keyword search}. We tried this for RTO, searching with the
term \emph{``rehabilitation''}. This naive approach does not work well, as it
misses many papers which contain the same stem but with a different
ending (such as ``rehabilitate'' or ``rehabilitator''). Moreover, it
retrieves many less relevant results (for example, those relating to
drug rehabilitation).

\begin{figure}
  \centering
  \includegraphics[scale=0.5]{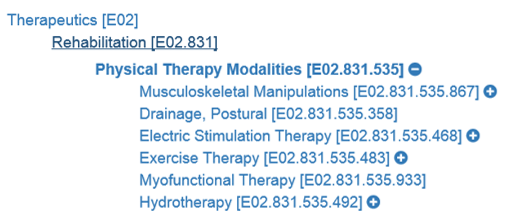}
  \caption{``Rehabilitation'' MeSH term with some of its narrower terms}
  \label{fig:mesh}
\end{figure}

Our next approach is to use MeSH or MAJR terms. PubMed's search engine
automatically searches the transitive closure of any MeSH term given,
therefore searches with \emph{``Rehabilitation''} will also search \emph{``Physical
Therapy Modalities''}, as can be seen in figure~\ref{fig:mesh}.

Clearly searching for \emph{``Rehabilitation''} as the MAJR term will produce
a result which is an exact subset of searching for the equivalent MeSH
term. In fact, the simple search approach automatically incorporates
MeSH search, as PubMed's search engine translates search terms to its
equivalent MeSH term if it exists. For example, the term
``Physiotherapy'' is translated to the ``Physical Therapy Modalities''
MeSH term.

MeSH search approach also runs the risk of missing papers which
have not been annotated at all, or have been annotated with
alternative terms from MeSH.

To address this latter problem, we have tried \textbf{query
  expansion}. Here, we expand the transitive closure of the MeSH term,
then add alternative endings manually. Sub-terms, more specifically
narrower terms, of ``rehabilitation'' were extracted using ``MeSH
SPARQL'' tool \footnote{MeSH SPARQL is available at
  https://id.nlm.nih.gov/mesh/query}. The following SPARQL query was
used:

\begin{lstlisting}[language=sparql,  basicstyle=\tiny, escapeinside=`']
PREFIX rdfs: <`\url {http://www.w3.org/2000/01/rdf-schema#}'>
PREFIX meshv: <`\url {http://id.nlm.nih.gov/mesh/vocab#}'>
PREFIX mesh: <`\url {http://id.nlm.nih.gov/mesh/}'>

SELECT ?label
FROM <`\url {http://id.nlm.nih.gov/mesh}'>
WHERE {
  ?term meshv:broaderDescriptor+ mesh:D012046 .
  ?term rdfs:label ?label .
}
\end{lstlisting}

We collected and filtered general terms used in other domains such as
\emph{``Yoga''} and \emph{``Massage''} by inspection. These are mostly
the ones without medical words such as rehabilitation or
therapy. Synonyms of the term \emph{``rehabilitation''} were defined
by consultation with a domain expert: specifically,
\emph{``restoration''} and \emph{``recovery''}. Multiple variations of
these words were determined manually using a dictionary. Variations of
the words \emph{``therapy''}, \emph{``rehabilitation''} include
\emph{``therapies''}, \emph{``therapist''} and
\emph{``rehabilitant''}.  rehabilitated, and were injected in the
query. Finally, the collected general terms were combined into a MeSH
approach query, the rest of the terms were combined into a query that
is disjunctive between noun phrases and their variants. For example,
the term \emph{``physical therapy''} was converted to:

\begin{lstlisting}[language=sql,  basicstyle=\tiny]
  Physical therapy OR Physical AND
   (therapy OR therapies OR therapist OR therapists OR therapeutic OR ...
  OR      rehabilitation OR rehabilitate OR rehabilitator OR ...
  OR      restoration OR restore OR ...
  OR      recovery OR ...)
\end{lstlisting}

The two queries were combined to form the expanded query.The result of
this approach subsumes the results of the two previous
approaches. Thus, this approach provides the most coverage. In fact,
we retrieved around 2.9 million MEDLINE records using the query expansion
approach. Table~\ref{Tab:01} shows the search terms for each approach
along with the number of retrieved records.

\begin{table}[h]
  \begin{tabular}{|c|c|c|}\hline
    Search Strategy  & Query Search Term(s) & Number of \\ & & retrieved records \\\hline
    Simple & Rehabilitation & 512,901 \\ \hline
    MeSH   & Rehabilitation [MeSH] & 258,541   \\\hline
    MAJR   & Rehabilitation [MAJR] & 156,038   \\\hline
    Query expansion  & The expanded query & 2,880,858 \\ 
                     & (as explained in the example)    & \\\hline
   \end{tabular}
   \caption{Search terms and the number of retrieved MEDLINE records for
     each of the three search approaches\label{Tab:01}}
 \end{table}

The query expansion search approach provides a significant increase in the
number of records. We tested the accuracy of the approach by random
selection of papers, followed by expert analysis to determine whether
the papers were in scope. Unfortunately, the accuracy of this approach
appears fairly low, with around 5\% of the papers considered in scope.

Finally, we have pioneered a \textbf{relative similarity}
measure. This builds on PubMed's existing article similarity score, and
allows us to define similarity to a set of articles. Retrieved
records are ranked with a relatively score which is calculated as follows:

\begin{align*}
  \label{eq:01}
  relativity \ score(a) = \frac {\# similar \ articles(a) \ that \ are  \ in \ s}{max(\# s, \# similar \ articles(a))}\\
  where \ s: \ seed \ set, \ a: \ article \ (i.e. \ MEDLINE \ record)
\end{align*}

From this equation, for a record to have a relativity score of
$1.0$, all of its similar records need to cover all of the records
in the seed set. In other words, a record can only have a relativity
score of $1.0$ if its set of similar records is equivalent to the
seed set. If it has a similar record that is not in the seed set or
if there is a record in the seed set that is not similar to it, the
relativity score will be less than $1.0$. Thus, for higher scores, a
record not only must be similar to more records in the seed set,
but also needs to have fewer similar records out of the seed set.

\begin{figure}[htpb]
  \centering
  \includegraphics[width=\linewidth, scale=0.6]{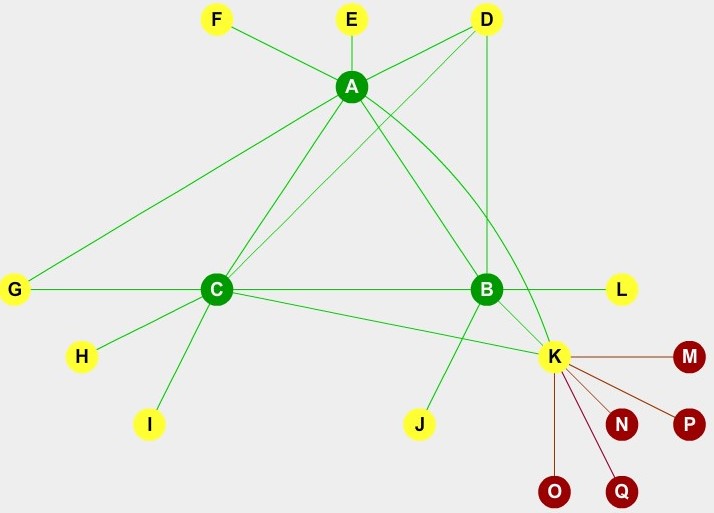}
  \caption{An example for relatively similar records approach. The
    green nodes represent the seed set of MEDLINE records
    (i.e. records), the yellow ones represent similar records, the
    red ones represent records similar to the yellow nodes records
    and the edges represent PubMed's similarity relation, which is
    symmetric.}
  \label{fig:relative}
\end{figure}

Figure~\ref{fig:relative} shows an example of this approach. There are
$3$ seed  MEDLINE records (i.e. records). The relativity score for the
node $D$ is $1.0$ , as all of its similar records are in the seed
set.  Below are some of the other records scores:

$$ relativity \ score(E) = \frac{1}{3}  $$
$$ relativity \ score(G) = \frac{2}{3}  $$
$$ relativity \ score(K) = \frac{3}{8}  $$

Although $K$, like $D$, is similar to all the records in the seed
set, unlike $D$, its score is lower than that of $G$ as it has more
similarity with other records out of the seed set. Records with
higher scores can be considered as more relatively similar to the seed
set. A significant advantage of this approach is that the result is
continuous and can be thresholded according to contain more or less
papers as required.

After achieving a maximal set of citations covering the topic, a
minimal accurate set was provided by a domain expert. The expert set
of articles was provided as an EndNote library file. We converted the
articles in the library file to PMIDs. We can test the coverage of
the maximal set by checking whether it subsumes the minimal set. In
fact, all of the articles provided by the expert were included in the
maximal set.

\begin{figure}[htpb]
  \centering
  \includegraphics[width=\linewidth]{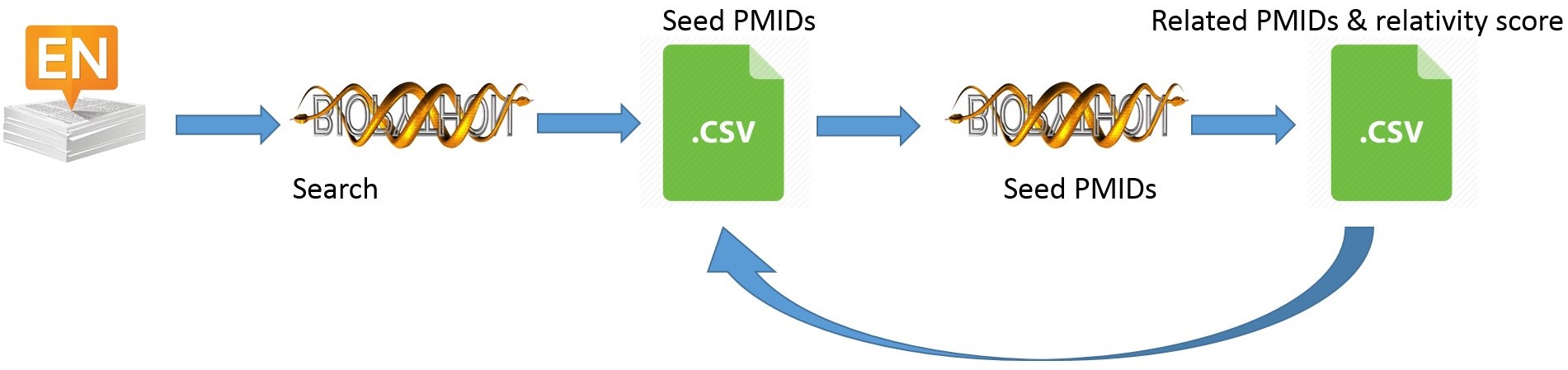}
  \caption{The process of retrieving a suitable corpus that is
    relatively similar to a seed corpus provided by the expert in
    EndNote library format.}
  \label{fig:process}
\end{figure}

Now, we can use this approach to retrieve relatively similar articles
from the experts seed set, i.e. the minimal set. The retrieved
articles that are not included in the maximal set are filtered to
restrict similar articles that are out of the maximal set's scope. The
expert, then, can set a threshold score to select the most related
articles. The articles above the threshold, or ones chosen by the
expert, can then be added to the seed set to perform the process
again. This process can be repeated iteratively with the help of the
expert until the results are satisfying or until they converge. The
choice of the threshold might partly depend on the required number of
retrieved articles, especially in the final stages. This process is
depicted in Figure~\ref{fig:process}.

\section{Conclusion}
\label{Conclusion}

In this article, we described four complementary search strategies to
retrieve an accurate and covering corpus of PubMed records for the
topic of rehabilitation. We use this approach to ensure that we have a
covering and unbiased corpus. Of the approaches tried, the simple
search and MeSH based strategies were too restrictive, the expanded
query too broad. To address these issues, we have developed a new
measure for paper similarity which enables us to select papers similar
to a group of papers. This approach enables us to threshold
arbitrarily and define for ourselves the ``Goldilocks'' zone.

The key advantage of this technique is that it requires relatively
little from the domain expert, beyond a set of references to
appropriate papers, something that most researchers will have through
their normal bibliography management facilities. Operationally, this
technique is also straight-forward as it works on PubMed similarity
(although it generalises to any similarity measure), and can operate
directly over PubMed's normal search facilities. This avoids the
necessity of performing bespoke analysis over the whole of PubMed
locally.

A significant advantage of this technique is that it works on PubMed
similarity (although it could work on any pair-wise similarity
metric), which makes it easy to perform. We can envisage perhaps
richer techniques that generalize the current over PubMed's similar
articles approach. However, until and unless these are directly
supported by PubMed, they would require warehousing PubMed locally.
For the next step, we plan to use this corpus to define a covering set
of terms for the Rehabilitation Therapy Ontology, using inverse
document frequency statitics that we have previously used to define
the scope of a minimum information model~\citep{lord2016minimum}.

We note that this approach is largely independent of domain. We do not
require a suitable MeSH term, or a pre-existing set of keywords that
can be used for querying. It raises the possibility of moving the
initial knowledge capture stage of ontology development away from
expert user groups and competency questions, toward an approach which
is more data-driven, embedding ontology development in the explosion
of interest in big data analytics that have characterised the last few
years.

\bibliographystyle{natbib}

\bibliography{2017_06_mohammad_icbo}

\end{document}